\documentclass[%
 reprint,
 amsmath,amssymb,
aps,
pra,
superscriptaddress,
longbibliography,
]{revtex4-2}
\usepackage{multirow}
\usepackage{graphicx}% Include figure files
\usepackage{dcolumn}% Align table columns on decimal point
\usepackage{bm}% bold math
\usepackage{mhchem}
\usepackage{braket}
\begin{document}

\preprint{APS/123-QED}
\title{Towards a Realistic Model for Cavity-Enhanced Atomic Frequency Comb Quantum Memories}% Force line breaks with \\
\author{Shahrzad Taherizadegan}
\affiliation{Department of Physics \& Astronomy, Institute for Quantum Science and Technology, University of Calgary, 2500 University Drive NW, Calgary, Alberta T2N 1N4, Canada}
\author{Jacob H. Davidson}
\affiliation{QuTech and Kavli Institute of Nanoscience, Delft University of Technology, 2600 GA Delft, The Netherlands} \thanks {Present Address: National Institute of Standards and Technology (NIST), Boulder, Colorado 80305, USA}
\author{Sourabh Kumar}
\affiliation{Department of Physics \& Astronomy, Institute for Quantum Science and Technology, University of Calgary, 2500 University Drive NW, Calgary, Alberta T2N 1N4, Canada}
\author{Daniel Oblak}
\affiliation{Department of Physics \& Astronomy, Institute for Quantum Science and Technology, University of Calgary, 2500 University Drive NW, Calgary, Alberta T2N 1N4, Canada}
\author{Christoph Simon}
\affiliation{Department of Physics \& Astronomy, Institute for Quantum Science and Technology, University of Calgary, 2500 University Drive NW, Calgary, Alberta T2N 1N4, Canada}

\begin{abstract}
Atomic frequency comb (AFC) quantum memory is a favorable protocol in long distance quantum communication. Putting the AFC inside an asymmetric optical cavity enhances the storage efficiency but makes the measurement of the comb properties challenging. We develop a theoretical model for cavity-enhanced AFC quantum memory that includes the effects of dispersion, and show a close alignment of the model with our own experimental results. Providing semi quantitative agreement for estimating the efficiency and a good description of how the efficiency changes as a function of detuning, it also captures certain qualitative features of the experimental reflectivity. For comparison, we show that a theoretical model without dispersion fails dramatically to predict the correct efficiencies. Our model is a step forward to accurately estimating the created comb properties, such as the optical depth inside the cavity, and so being able to make precise predictions of the performance of the prepared cavity-enhanced AFC quantum memory. 
\end{abstract}

\maketitle

%\tableofcontents

\section{Introduction}
Optical quantum memory with the ability to store and recall on-demand quantum states of light with high efficiency and fidelity \cite{simon2010quantum, heshami2016quantum, ma2020optical} is one of the essential elements for long-distance quantum communication based on quantum repeaters \cite{briegel1998quantum,duan2001long}. It also has several applications in linear-optical quantum computation, single-photon detection, quantum metrology, and tests of the foundations of quantum physics \cite{bussieres2013prospective}. 
\\\indent AFC quantum memory \cite{afzelius2009multimode,de2008solid} is a promising candidate in quantum repeater applications because of the capability to simultaneously store and read out multiple temporal and spectral modes that could enhance the performance of the quantum repeater via faster entanglement generation \cite{simon2007quantum}. Also, as opposed to other quantum memory protocols, with the AFC technique, the number of temporal modes stored in a sample is independent of the optical depth of the storage medium. 
\\ \indent To implement AFC quantum memory, rare-earth-ion doped crystals are particularly suitable due to the long coherence times of their optical
4f-4f transitions at cryogenic temperature\cite{guo2023rare}. Long optical storage time has been demonstrated in rare-earth ion-doped crystal-based AFC quantum memory by using dynamical decoupling techniques to increase the spin coherence time \cite{Holzäpfel_2020, Ma2021, Ortu2022}. There have been attempts to improve the efficiency of the AFC quantum memory by enhancing the preparation procedure of the AFC including the comb shape optimization in Tm:YAG crystal \cite{bonarota2010efficiency} and some improvement in the efficiency has been demonstrated compared to the initial preparation procedure of the AFC \cite{chaneliere2010efficient,de2008solid}.
\\\indent To achieve high efficiency in quantum memories, a large optical depth in the storage material is typically needed. However, in practice, simultaneously achieving high optical depth and long coherence times is difficult \cite{tittel2010photon}. This trade-off is exemplified by the observation that the coherence time is usually inversely dependent on the doping concentration. So, to have high coherence times, it is preferable to work with lower optical depth (OD) rare-earth doped crystals. To overcome this limitation, it was proposed to incorporate the storage medium in an asymmetric optical cavity. By applying the impedance matching condition, unit efficiency can, in principle, be obtained with an effective optical depth of only one \cite{afzelius2010impedance, PhysRevA.82.022311}. In this situation, the memory efficiency is only limited by intrinsic atomic dephasing. \\\indent So far, several experiments have been carried out based on the impedance matching proposal \cite{afzelius2010impedance, PhysRevA.82.022311} using the AFC technique \cite{sabooni2013efficient, jobez2014cavity,akhmedzhanov2016cavity, davidson2020improved}. To date, the best efficiencies obtained are 62\% \cite{duranti2023efficient} for the storage time of $2 \, \mu s$, 56\% \cite{sabooni2013efficient} for the storage time of $1.1 \, \mu s$ and 53\% \cite{jobez2014cavity} for the storage time of $2 \, \mu s$. Based on the experimental results, between 12-20 fold enhancement in the storage efficiency has been obtained by putting the AFC memory in an impedance-matched optical cavity compared to the no cavity case. However, it has been experimentally difficult to measure the AFC properties, e.g., optical depth within the impedance matched cavity \cite{sabooni2013efficient,akhmedzhanov2016cavity,davidson2020improved}. 
The fact that no detailed general theoretical model for cavity-enhanced AFC quantum memory exists, makes it difficult to compare experiments to theory, and hence also to infer the system parameters from experiments. The original proposal \cite{afzelius2010impedance,PhysRevA.82.022311} assumed a memory bandwidth significantly smaller than the cavity bandwidth to satisfy the resonance condition, and so did not discuss the combs that are created at frequencies detuned from the cavity resonance. Also, the background absorption due to imperfect optical pumping used to create the comb shape was ignored. Although, authors in Ref. \cite{bonarota2012atomic, berman2021pulsed} have investigated the role of dispersion in AFC as a protocol closely associated to the slow-light-based storage protocols, so far there has not been a theoretical model for cavity-enhanced AFC quantum memory that includes the dispersion originating from the absorption engineering of ions to create the comb inside the cavity. 
\\\indent Here we develop a more general model that addresses all of these points. We extend the impedance-matched model beyond the resonance condition by including the round-trip phase shifts of light as travelling inside the cavity in the initial proposal \cite{afzelius2010impedance} making it valid for any AFC bandwidth with a background absorption, and created at any detuning with respect to the cavity resonance. We show that, including dispersion, our developed model closely aligns with our own experimental results, and in particular enables prediction of the experimental memory efficiency at any detuning with respect to the cavity resonance. The paper is organized as follows: In sections \ref{sec:theory} and \ref{expr} the theoretical model and the experiment performed are discussed. In section \ref{Results} the method to predict the cavity-enhanced AFC quantum memory reflectivity and efficiency is explained. Furthermore, we show results from the theory and compare them to the experimental data. In section \ref{comparison} we elaborate on a comparison to a theoretical model without dispersion, and the conclusion and outlook are given in section \ref{conclusion}.

\section{Theoretical model for cavity-enhanced atomic frequency comb quantum memory}
\label{sec:theory}
\begin{figure}[h!]
\centering
\includegraphics[scale=0.4]{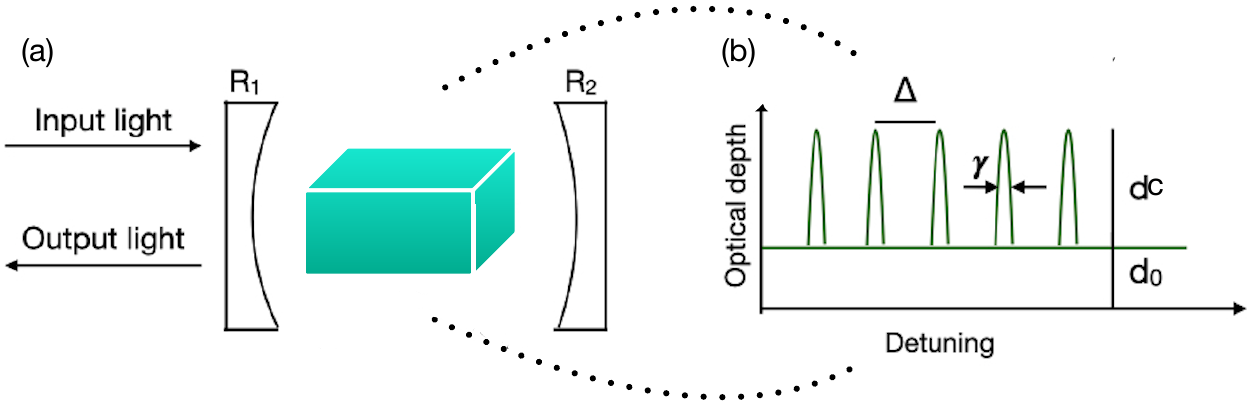}
\caption{\textbf{(a)} Crystal inside a general asymmetric cavity with mirror reflectivities $R_1$ and $R_2$ where $R_1 < R_2 \approx 1$. \textbf{(b)} The natural inhomogeneously broadened absorption profile of the medium (Tm:YAG in this paper) is shaped into a comb structure where the peaks are separated by $\Delta$ and have a width (FWHM) of $\gamma$ as shown on the right. The finesse of the comb is defined as $\mathrm{F=\frac{\Delta}{\gamma}}$. Note FWHM peak width $\gamma$ is related to the peak width $\tilde{\gamma}$ through $\gamma = \sqrt{8 \ln 2} \, \tilde{\gamma}$. Due to the comb structure, there will be a coherent re-emission after a time $\frac{1}{\Delta}$.}  
\label{My drawing}
\end{figure}

An AFC quantum memory consists of an ensemble of inhomogeneously broadened atoms engineered as periodically comb-like peaks in frequency domain (see figure \ref{My drawing}) which is realized through frequency-selective optical pumping of atoms from the ground state to a metastable state, e.g., a hyperfine state. A resonant input pulse with the bandwidth that matches the comb bandwidth is collectively absorbed. After absorption of the light, the collective excitation initially starts dephasing; however, after a time $t=1/\Delta$, ($\Delta$ is the spectral distance between the peaks in $\mathrm{Hz}$), due to the periodic structure of the comb, the excitation will be in phase again which leads to the echo pulse. 
\\\indent We consider an AFC quantum memory with a background absorption inside a general asymmetric cavity with mirror reflectivities $R_1$ and $R_2$ where $R_1 < R_2 \approx 1$ (see figure \ref{My drawing}) and apply the ”sum over all round-trips” approach of a Fabry-Perot cavity. We include absorption factors both for the frequency comb and the background due to additional background absorption to obtain the reflected intensity from the cavity. Furthermore, to extend the impedance-matched proposal for off-resonant combs with an arbitrary bandwidth, we include the phases associated with each path in the sum over all round trips (compared to the on resonance equation $(11)$ in Ref. \cite{afzelius2010impedance}). The reflected amplitude $(E_{out}/E_{in})$ from the cavity \cite{pollnau2020spectral, siegman1986lasers} can be written as 
\begin{equation}
\frac{E_{out}}{E_{in}}= \frac{-\sqrt{R_1} + \sqrt{R_2}  e^{-d(\nu)} e^{-i \Phi(\nu)}}{1-\sqrt{R_1 R_2} e^{-d(\nu)} e^{-i \Phi(\nu)}}\;\; ,
\label{2}
\end{equation}
where $L$ is the length of the crystal, $d(\nu) = \alpha(\nu) \, L$ is the optical depth, and $\alpha(\nu)$ is the absorption coefficient. In addition, the total round-trip phase is $\Phi = 2KL$, where $K = 2\pi \frac{n(\nu)}{\lambda}$ is the wavenumber, and $n(\nu)$ is the real refractive index of the matter inside the cavity. 
$\Phi(\nu)$ can be written as
\begin{equation}
\Phi(\nu) = 4\pi\frac{n(\nu)\, \nu L}{c}\;\; ,
\label{phi}
\end{equation}
where $c$ is the speed of light in the vacuum. 

\begin{figure}[h]
\centering
\includegraphics[scale=0.2]{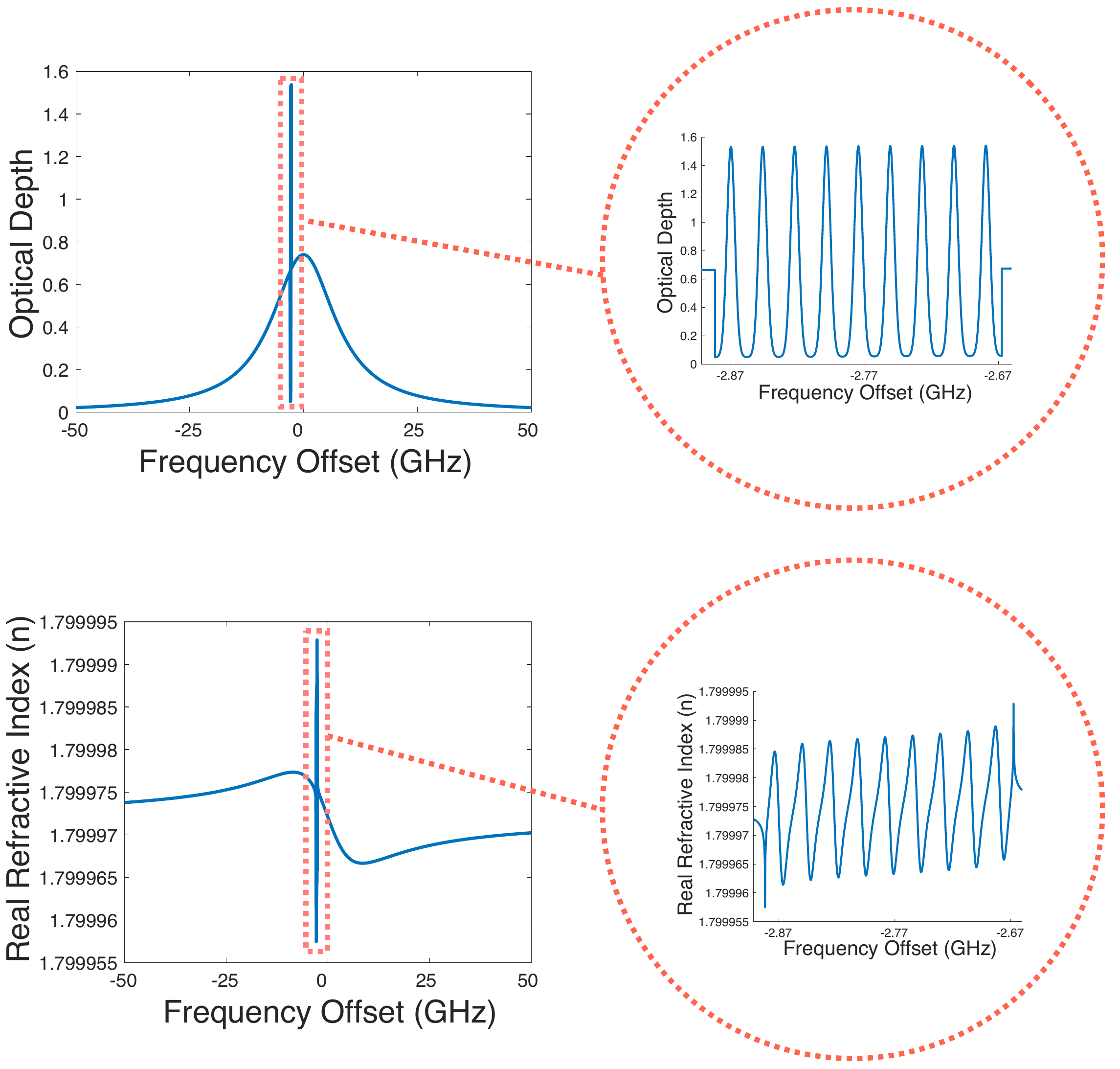}
\caption{\textbf{(a)} The natural inhomogeneously broadened optical depth of Tm:YAG crystal with the comb-shape engineered optical depth for the comb at $-2.7720 \, \textrm{GHz}$ detuning with respect to the center of the inhomogeneous broadening embedded into that. (\textbf{Inset}) The created comb-shaped optical depth over the comb bandwidth. \textbf{(b)} The calculated real refractive index $n(\nu)$ over the entire frequency range for the optical depth shown in (a). (\textbf{Inset}) $n(\nu)$ over the comb width.}
\label{ODRealn}
\end{figure}

\indent Equation \eqref{2} for the reflected amplitude can be interpreted as the response function of the cavity. We will refer to both terms interchangeably. Since $E_{out}$ contains terms which depend on the optical depth and the phases associated with each path in the sum over all round-trips, the reflectivity of the cavity will become frequency dependent over the comb bandwidth. Equation \eqref{2} is a general version of equation $(11)$ in Ref. \cite{afzelius2010impedance} and can be applied to both on-resonance and off-resonance conditions. 
In section \ref{Results}, we use equation \eqref{2} to simulate the reflectivity of the cavity and compare it with the experimental data.
\\\indent We assume an AFC with an engineered optical depth $d(\nu)$ of a series of Gaussian peaks with identical amplitudes ($d_{c}$), spacing ($\Delta$), width ($\tilde{\gamma}$), and a constant $d_0$ which is the optical depth associated with the background absorption. The AFC optical depth $d(\nu)$ can be written as 
\begin{equation}
d(\nu) = \sum_{k=1}^{9} \,d_c\, e^{-{\frac{(\nu-b_{k})^2}{2\tilde{\gamma}^2}}} + d_0 \; \; ,
\label{d}
\end{equation}
where the number $9$ in the summation is the number of created teeth in the experiment, $b_{k}$ is the center of each Gaussian function (each AFC tooth) and $b_{k} - b_{k-1} = \Delta$ (see figure \ref{My drawing}). The optical depth of the crystal defines the initial $\alpha(\nu)$, and optical pumping to define the comb reshapes the natural absorption profile. The resulting profile can be described using $d_0$ for residual unaffected absorption, and $d(\nu)$ for the total absorption corresponding to the newly shaped comb features. 
\\\indent By including the dispersion originating from the atomic absorption in the model there is a frequency-dependent refractive index $n(\nu)$ in the phase $\Phi$. The complex refractive index $\tilde{n}(\nu)$ can be written as $\tilde{n}(\nu) = n(\nu)+ik(\nu)$. The real part of the complex refractive index $n(\nu)$ is responsible for the change in the phase of light (see equation \eqref{phi}). The imaginary part of the complex refractive index $k(\nu)$ is related to the absorption coefficient as,
\begin{equation}
k(\nu) = \frac{\alpha(\nu)\,c}{4\pi\,\nu}\; \; . 
\end{equation}
The real and imaginary parts of the complex refractive index $\tilde{n}(\nu)$ are connected through the Kramers-Kronig relations, through the real $\chi_r(\nu)$ and imaginary $\chi_i(\nu)$ parts of the susceptibility of a medium \cite{kronig1926theory, kramers1927diffusion, toll1956causality, lucarini2005kramers, Saleh1991}
\begin{equation*}
\chi_r(\nu)= \frac{2}{\pi} \mathcal{P} \int_0^\infty \frac{\nu' \chi_i(\nu')}{\nu'^2 - \nu^2} d\nu'  \; \; ,
\end{equation*}
\begin{equation}
\chi_i(\nu)= \frac{2}{\pi} \mathcal{P} \int_0^\infty \frac{\nu \chi_r(\nu')}{\nu^2 - \nu'^2} d\nu'          
\label{5}
\end{equation}
where $\mathcal{P}$ symbolizes the Cauchy principal value. Note that in these equations $\nu$ is the absolute frequency, whereas we show the detuning in most of the rest of the paper. Thus, a frequency-dependent absorption coefficient results in a frequency-dependent $k(\nu)$ and it affects the real refractive index $n(\nu)$. Using the Kramers-Kronig relations (equation \eqref{5}) we can write the the real part of the complex refractive index $n(\nu)$ as
\begin{equation}
n(\nu) = n + \frac{2}{\pi} \mathcal{P} \int_0^\infty \frac{\nu' k(\nu')}{\nu'^2 - \nu^2} d\nu' \; \; ,
\label{6}
\end{equation}
$n$ is the constant real refractive index of the host crystal ($\mathrm{YAG}$ in this paper) \cite{zelmon1998refractive,hrabovsky2021optical}. Putting the obtained $n(\nu)$ for the comb in equation \eqref{phi}, the change in phase of light $\Phi$ is calculated and one can write the equation for $\Phi$ as 
\begin{equation}
\Phi(\nu) = 4\pi \frac{\nu L}{c}\,\left(n + \frac{2}{\pi} \mathcal{P} \int_0^\infty \frac{\nu' k(\nu')}{\nu'^2 - \nu^2} d\nu'\right)
\label{7}
\end{equation}
Note that to calculate the real refractive index $n(\nu)$ at each frequency the absorption coefficient must be defined over the entire frequency range. Figure \ref{ODRealn}\,\textbf{a} shows an example of the employed optical depth for a comb created close to the cavity resonance. 
Employing such an absorption coefficient allows us to calculate the real refractive index $n(\nu)$ over the comb. Figure \ref{ODRealn}\,\textbf{b} shows the calculated real refractive index $n(\nu)$ for the comb shown in figure \ref{ODRealn}\,\textbf{a}. Using equation \eqref{7} for $\Phi$, one can calculate the reflectivity of the cavity-enhanced AFC quantum memory as a function of frequency which will be done in section \ref{Results}. 

\begin{figure*}
\includegraphics[width=0.95\textwidth]{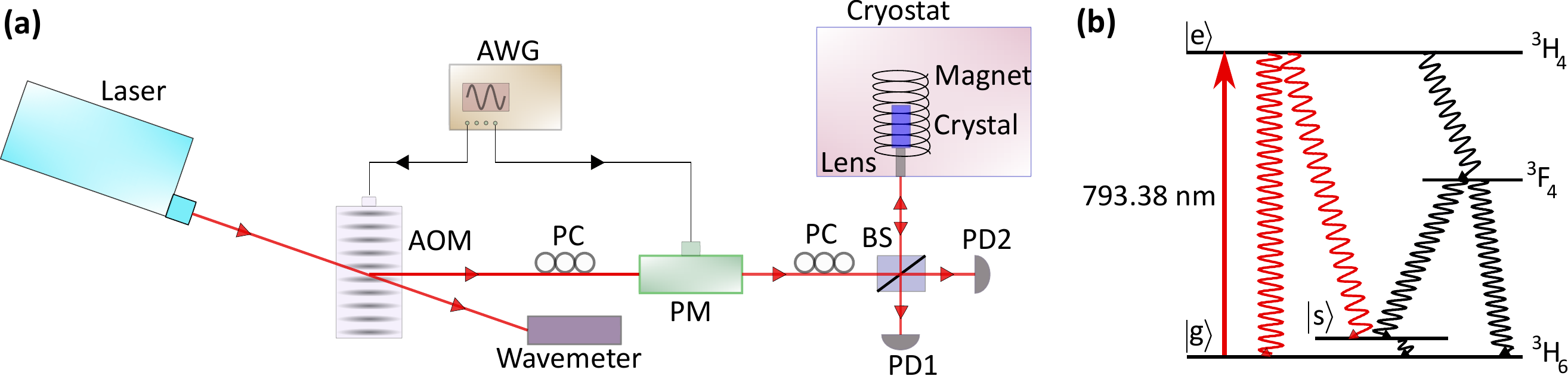}
\caption{\label{fig:wide} \textbf{(a)} Schematic of the experimental setup. A tunable external cavity diode laser serves as the light source. The first order of the acousto-optic modulator (AOM) goes through a polarization controller (PC) before passing an electro-optic phase modulator (PM). An arbitrary waveform generator (AWG) sends electrical signals to drive the AOM and the PM and control the experimental sequence. The 50/50 beamsplitter ensures that the same optical fiber transmits light to the crystal and collects the reflected light. The photodetector PD1 detects the back-reflected light from the crystal-cavity whereas detector PD2 monitors the input laser light intensity. The sample is kept inside an adiabatic demagnetization refrigerator (ADR) and another superconducting magnet placed inside the cryostat applies a DC magnetic field to the crystal sample. \textbf{(b)} Energy level diagram of \ce{Tm^{3+}} ion. For optical pumping, the laser is tuned to the $\ket{g} \rightarrow \ket{e}$ transition. Under the application of an external magnetic field, the ground state splits into Zeeman levels $\ket{g}$ and $\ket{s}$. The latter serves as the auxiliary level for the formation of the atomic frequency comb (AFC). The echo after the storage time is emitted at the $\ket{g} \rightarrow \ket{e}$ transition frequency.} 
\label{exprset}
\end{figure*}

\section{Experiment}
\label{expr}
We have performed our experiments in  a $0.1\%$ thulium-doped $\mathrm{Y_3 Al_5 O_{12}}$ $\textrm{(Tm:YAG)}$ crystal cavity, the same as used in reference \cite{davidson2020improved}. The crystal is approximately $4\,\textrm{mm}$ long and the AFCs are generated at a cryogenic temperature of $1.5$ K. The two ends of the crystal surface are reflection coated with approximate reflectivities $R_1 = 40 \%$ on the front end and $R_2 = 99 \%$ on the back end \cite{davidson2020improved} to satisfy the impedance matching condition. As a result, the created asymmetric optical cavity has a finesse close to $7$ and free spectral range close to $20$ GHz. 
\\\indent Figure \ref{exprset}\,\textbf{a} shows the experimental setup. Light near 793 nm from an external cavity diode laser (ECDL) passes through an acousto-optic modulator (AOM) driven at 400 MHz. While the zeroth order of the AOM is monitored on a wavemeter to measure the wavelength of the laser, the first diffracted order is directed to the crystal in the cryostat after passing through a phase modulator (PM) and a 50/50 beamsplitter. The energy level structure of \ce{Tm^{3+}} ions is shown in figure \ref{exprset}\,\textbf{b}. Under the application of an external magnetic field, the ground state \ce{^{3}H_{6}} splits into the Zeeman levels $\ket{g}$ and $\ket{s}$. The phase modulator is used to carve out the combs by selectively pumping ions in the frequency domain from the ground state $\ket{g}$ to the spin state $\ket{s}$ \cite{davidson2020improved}, mediated by the excited state $\ket{e}$. The 50/50 beamsplitter is employed to be able to send and receive photons by back-reflection from the cavity using the same optical fiber. The optical fiber is terminated in a ferrule, which is placed in front of a gradient index (GRIN) lens. Both the ferrule and lens are inserted in a tight-fitting 50 mm-long capillary, which is fixed at one end to the crystal mount and at the other to a xyz nano-positioning stage. In this way, movement of the nano-positoner will change the angle of the capillary, and, thus, the enclosed fibre and GRIN lens, to the crystal surface allowing the input mode to be aligned to the cavity mode.
\\\indent AFC combs are created at varying frequencies across the cavity spectrum by tuning the ECDL frequency. For measurements of the cavity reflection spectrum, the ECDL frequency can also be continuously swept across nearly 50 GHz. The strength of the external magnetic field controls the Zeeman splitting between the spin levels $\ket{g}$ and $\ket{s}$. We choose a magnetic field value close to 500 Gauss such that during the optical pumping in the AFC preparation, the atoms from the AFC troughs are almost all moved to the AFC peaks, increasing their optical depth nearly by a factor of two \cite{davidson2020improved}. This efficient AFC preparation ensures that almost all the atoms participate in the absorption and re-emission process. An arbitrary waveform generator (AWG) drives the AOM and the PM to create the experimental sequence for optical pumping and input pulse generation. After 50 ms optical pumping and 5 ms wait time, we send in narrow-band input pulses generated by the AOM. The back-reflected light, after passing through the beamsplitter, is collected on one of the photodetectors (PD1 in figure \ref{exprset}\,\textbf{a}). We tune the laser to different wavelengths and record the AFC scans at each wavelength. The scans are obtained with the help of a phase modulator, which generates a linear chirp spanning the whole AFC width. The frequency spanned is $200 \,\mathrm{MHz}$ in $100 \,\mathrm{\mu s}$. The photodetector PD1 detects the chirped light after interacting with the AFC profiles and the AFC scans are obtained on an oscilloscope connected to PD1. We repeat the experimental sequence $100$ times with a $200 \, \mathrm{ms}$ interval between each repetition to reduce the experimental noise and generate a steady AFC. We take the average of these $100$ scans on the oscilloscope to obtain the effective AFC trace. After the AFC trace is collected at a particular wavelength, we send an input pulse centered at that wavelength. We observe the unabsorbed part of the pulse and the echoes emitted on the oscilloscope connected to PD1. Similar to the AFC scans, we repeat the experimental sequence $100$ times and take the average to obtain the effective traces for the pulses and the echoes. To obtain the input electric field $(E_{in})$ in Equation \eqref{2}, we go to an off-resonant frequency, greater than $200 \,\mathrm{GHz}$ away from the center of the inhomogeneous broadening. Since almost no atoms interact with the incident light at this wavelength, the AFC trace is close to a flat line. Also, virtually no part of the incident pulse is absorbed at this wavelength and no echoes are emitted. The unabsorbed part of the incident pulse is now taken as the input pulse for calculating the memory efficiencies. 
We monitor the fluctuations in the input power of the laser as we tune its wavelength by connecting the free port of the beamsplitter to another photodetector (PD2) as shown in figure \ref{exprset}\,\textbf{a}. We calculate the experimental efficiencies by dividing the area of the first echo at each wavelength by the area of the input pulse. The maximum experimental efficiency is obtained for the comb close to the cavity resonance frequency and at the detuning of $-2.7720\, \textrm{GHz}$ from the resonance frequency of the atoms (all the detunings in the present manuscript are with respect to the center frequency of the inhomogeneously broadened atomic transition, unless otherwise mentioned).   
\\\indent Using the theory discussed in section \ref{sec:theory} and this experimental setup we are able to extract some relevant features of our memories and verify the performance of our new model in the following section.

\section{\label{Results} Results and Discussion}
In this section, we show and analyze the experimental results. First, equation \eqref{2} is fitted with the experimental data for the reflectivity of the crystal cavity with no comb carved into it and the values of the crystal cavity properties are obtained as the fitting parameters in section \ref{ref nocomb}. Note that equation \eqref{2} is for the reflected amplitude and not the reflected power (${{\lvert \mathrm{reflected~amplitude} \rvert}^2 = \mathrm{reflected~power}}$). Next, the crystal cavity with a spectral-shaped AFC is considered. By fitting the model to the measurement results for the reflectivity and using the obtained fitting parameters for the crystal cavity properties from the previous step, the shape and optical depth of the created combs are extracted as shown in section \ref{ref comb}. It is worth to mention the fitting codes are all written in Matlab \cite{Lucarini2023}, and the curve fitting and optimization toolboxes are applied. Finally, the obtained comb properties from the fitting are used to predict the efficiencies of the created cavity AFC quantum memories at various frequency offsets across the cavity profile in section \ref{mem efficiency}, and the predicted memory efficiencies are compared to the experimental results. 

\subsection{Cavity reflectivity without comb}
\label{ref nocomb}
\begin{figure}[h]
\centering
\includegraphics[scale=0.5]{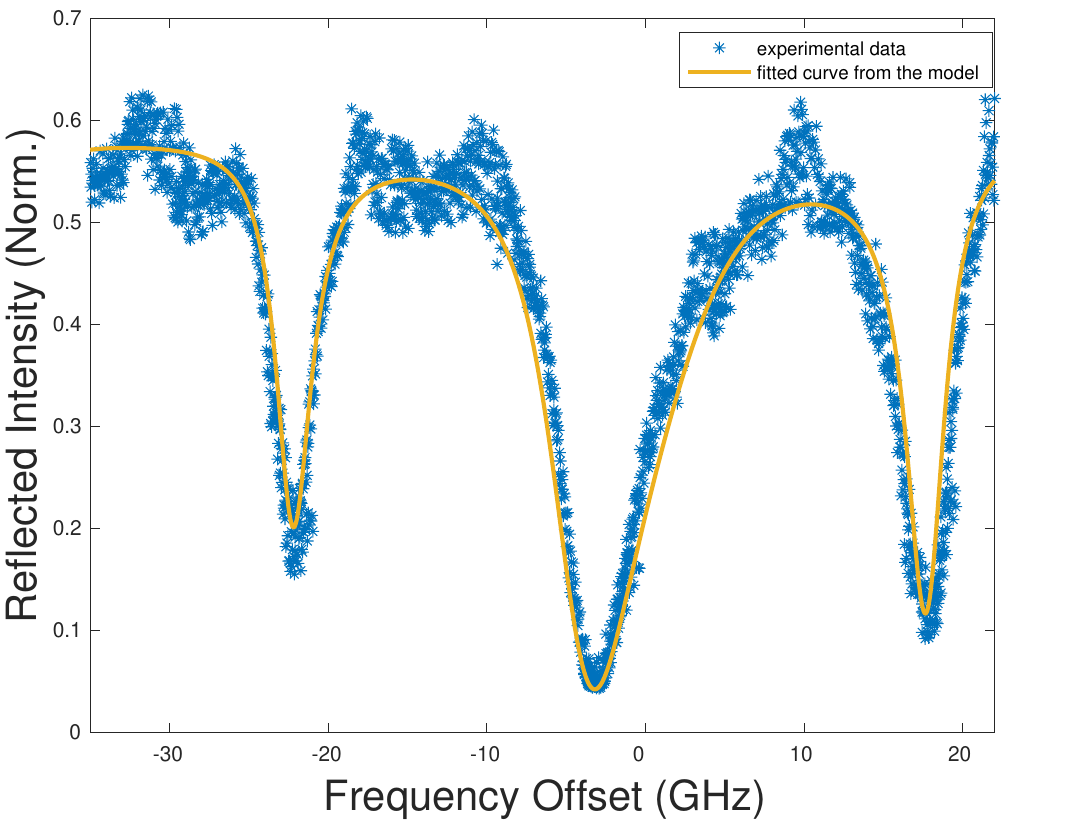}
\caption{Experimental cavity reflectivity and the fitted curve from the model.} 
\label{cavityref}
\end{figure}

The reflected power at different frequencies for the crystal cavity with no AFC carved into it can be obtained using equation \eqref{2}. The values of the crystal cavity properties are obtained as the fitting parameters from fitting equation \eqref{2} in the model with the experimental normalized reflected power for the
$\textrm{Tm:YAG}$ crystal cavity with no comb engineered into it measured by \cite{davidson2020improved}. The real refractive index $n(\nu)$ in equation \eqref{phi} is frequency dependent and can be obtained using equation \eqref{6} (section \ref{sec:theory}). The inhomogeneously broadened absorption spectrum of $\textrm{Tm}$ in $\textrm{YAG}$ is a Lorentzian function centered at $\mathrm{\nu_0} = 377, 868 \, \mathrm{GHz}$, with a FWHM of $\Gamma_{in} = 17 \, \mathrm{GHz}$ \cite{PhysRevB.94.205133}. Figure \ref{cavityref} shows the data for the reflectivity of the cavity at different frequencies and the fitted curve to the data. We see three cavity modes in the figure. Near impedance-matching occurs at the frequency offset of $-3.19\,\mathrm{GHz}$. It should be noted that the impedance-matched point is close to the perfect impedance-matching: theoretically $100\%$ of the input light should be absorbed at the impedance-matched frequency (zero reflection); however, the experimental value got close to $95\%$ absorption. The obtained impedance-matched frequency for the fitted curve is in good agreement with the experimental value of $-3.4\, \pm \, 0.5 \, \textrm{GHz}$ where the uncertainty in the measured value comes from the wavemeter's smallest significant digit. The maximum value of the Lorentzian function of the absorption spectrum is also set as a fitting parameter. This takes into account possible variations in crystal growth and sections of the boule from which our sample was cut. Although the experimental cavity reflection data (figure \ref{cavityref}) is measured through a $50/50$ beam splitter, the measured maximum reflectivity value of slightly more than $0.5$ in figure \ref{cavityref} implies the employed beam splitter was not exactly a $50/50$ one. To account for that there is a factor of $s$ as a fitting parameter so that (${\mathrm{reflected~power~/~s}}$) is fitted to the experiment, while $s=2$ is for an ideal $50/50$ beam splitter. The obtained values for the crystal cavity parameters are shown in table \ref{table:1}. Also, the obtained peak absorption coefficient (pac) is in good agreement with the measured value of between $1.9\, \mathrm{cm^{-1}}$ to $2.3\, \mathrm{cm^{-1}}$ \cite{liu2006spectroscopic, PhysRevB.94.205133}.
\\\indent In the next step the obtained fitting parameters for the crystal cavity properties are used to analyze the experimental reflectivity of the cavity-enhanced AFC quantum memory. 

\begin{table}[h!]
\caption{Obtained parameters for crystal cavity properties from cavity reflectivity without comb (see figure \ref{cavityref}).}
\label{table:1}
\begin{center}
\scalebox{0.85}{
\begin{tabular}{||c|c||}
\hline \hline
\multicolumn{2}{||c||}{Crystal Cavity Properties} \\
\hline \hline
pac\footnotemark[1]$(\mathrm{cm^{-1}})$ & 1.70 \\
\hline
$r_1=\sqrt{R_1}$ & 0.6927 \\ 
\hline
$r_2=\sqrt{R_2}$ & 0.9999 \\
\hline
n & 1.799972 \\
\hline
L $(\textrm{cm})$ & 0.4350 \\
\hline
s & 1.7142 \\
\hline \hline
\end{tabular}
}
\end{center}
\footnotetext[1]{peak absorption coefficient}
\end{table} 

\begin{figure*}[t]
\includegraphics[scale=0.385]{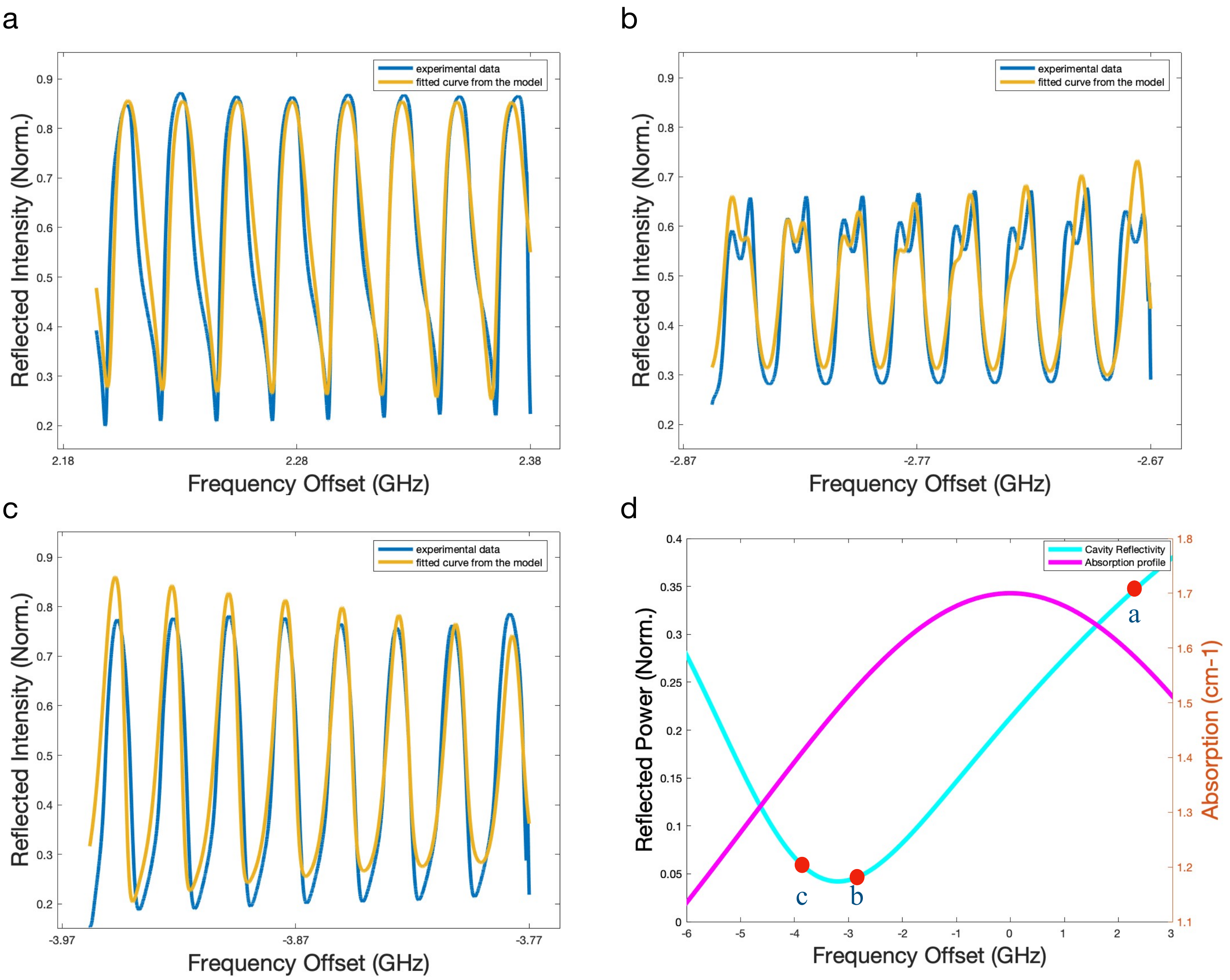}
\caption{Experimental cavity-enhanced AFC quantum memory reflectivity and the fitted curves for three combs created at three critical regions across the cavity features. (\textbf{a}) shows a comb far from the impedance matched frequency (at the detuning of $2.2765 \, \textrm{GHz}$). (\textbf{b}) shows a comb close to the impedance matched frequency (at the detuning of $-2.7720\, \textrm{GHz}$). (\textbf{c}) shows a comb in an intermediate regime (at the detuning of $-3.8675\,\textrm{GHz}$). (\textbf{d}) The inhomogeneously broadened absorption spectrum of $\textrm{(Tm:YAG)}$ crystal and the reflectivity of the $\textrm{(Tm:YAG)}$ crystal cavity over a smaller range (the range of the created combs). The locations of the above combs are labeled \textbf{a}, \textbf{b}, and \textbf{c} corresponding to the combs in subfigures \textbf{a}, \textbf{b}, and \textbf{c}.}
\label{combsall}
\end{figure*}

\begin{table*}
\caption{Obtained comb parameters for cavity-enhanced AFC Quantum memories created at different detunings across the cavity profile (see figure \ref{combsall}).}
\label{table2}
\begin{center}
\scalebox{0.89}{
\begin{tabular}{||c||c||c||c||}
\hline \hline
Combs' Properties &\multicolumn{1}{|c||}{Comb \textbf{a}} & \multicolumn{1}{|c||}{Comb \textbf{b}} & \multicolumn{1}{|c||}{Comb \textbf{c}} \\
\hline \hline 
$d_{c}$ & 1.5260 & 1.4867 & 1.4261 \\ 
$\Delta \, (\textrm{MHz})$ & 23.4598 & 23.8160 & 24.3382 \\ $\tilde{\gamma}\, (\textrm{MHz})$ & 3.6063 & 2.9755 & 3.4462 \\ $d_0$ & 0.2008 & 0.0526 & 0.0254 \\
\hline \hline
\end{tabular}
}
\end{center}
\end{table*}

\subsection{Cavity reflectivity with comb}
\label{ref comb}
In this section the experimental data for the reflected power of the cavity-enhanced AFC quantum memories created at different detunings are analyzed to extract the combs' features. As stated in the section \ref{sec:theory} it is assumed the AFC is a series of Gaussian functions with the same amplitude ($d_{c}$), spacing ($\Delta$), width ($\tilde{\gamma}$), and with a constant background absorption $d_0$ (see equation \eqref{d}). We use equation \eqref{7} to calculate the phase $\Phi$. Then equation \eqref{2} is fitted to the experimental data for the reflected power of each of the memories at different wavelengths and the comb parameters $d_{c}$, $\Delta$, $\tilde{\gamma}$, and $d_0$ are extracted as the fitting parameters. The fitted curves to the experimental reflected power using the model equations are shown in figure \ref{combsall} and the extracted fitting parameters are shown in table \ref{table2} for three AFCs created at three critical regions across the cavity features (see figure \ref{combsall}\,\textbf{d}). The first one is the AFC created far from the cavity resonance at the detuning of $2.2765 \, \textrm{GHz}$ with respect to the resonance frequency of the atoms (see figure \ref{combsall}\,\textbf{a}). The second one is close to the impedance-matched frequency at the detuning of $-2.7720\, \textrm{GHz}$ (see figure \ref{combsall}\,\textbf{b}). The third comb is in an intermediate regime, and on the other side of the cavity resonance at the detuning of $-3.8675 \,\textrm{GHz}$ (see figure \ref{combsall}\,\textbf{c}). 
\\\indent Figures \textbf{(a)}, and \textbf{(c)} in \ref{combsall} show there is a skewness at the bottom of the experimental data for the reflectivity of the combs created at $2.2765 \, \textrm{GHz}$ and $-3.8675 \, \textrm{GHz}$ which is captured in the fitted curves obtained from the model. In figure \ref{combsall} for the shown combs, the skewness at the bottom of the reflected power data changes direction as one moves from one side to the other side of the cavity resonance which is also true for all the other combs that are not shown, and is captured by the model. 
Also, for the comb created at $-3.8675 \, \textrm{GHz}$ detuning (see figure \ref{combsall} c) there is an asymmetry around its lowest point (skewness) at the bottom of the experimental data for all of the teeth in the comb captured to some extent in the model (for most of the teeth in the comb). Moreover, note the shapes at the top of the comb created close to the impedance-matched point at $-2.7720 \, \textrm{GHz}$ frequency offset, and the fact that the maximum of the experimental reflectivity is lower as compared to the other two combs. As one can see in figure \ref{combsall}\,\textbf{b}, relatively similar features are obtained by the model.  
\\\indent Discussing the obtained fitting 
parameters for the comb properties in the model, the spacing ($\Delta$) between the teeth which is related to the storage time of the AFC quantum memory is very similar amongst the created AFCs as is expected, and approximately the same as the model. Considering the storage time of $42 \, \textrm{ns}$ for the memories, one can obtain $\Delta = 23.81 \, \textrm{MHz}$ which is in good agreement with the extracted $\Delta$ values in table \ref{table2}. Thus, the challenges to understand the comb features are mainly focused on achieving the true values for the atomic absorption and the width of the related combs. The obtained value of $\tilde{\gamma}$ in the model for each comb shown in figure \ref{combsall} doesn’t change a lot within different combs resulting in nearly the same values for the finesse of different combs as expected from the experiment. Considering the pumping procedure (section \ref{expr}) and the inhomogeneous broadening in the absorption spectrum of $\textrm{Tm:YAG}$ crystal, we expect somewhat lower $d_{c}$ values compared to the ones obtained from the model. This leads to a difference between the theoretical and experimental efficiency which is discussed in the next section.

\subsection{Memory Efficiency}
\label{mem efficiency}
\begin{figure}[t]
\centering
\includegraphics[scale=0.45]{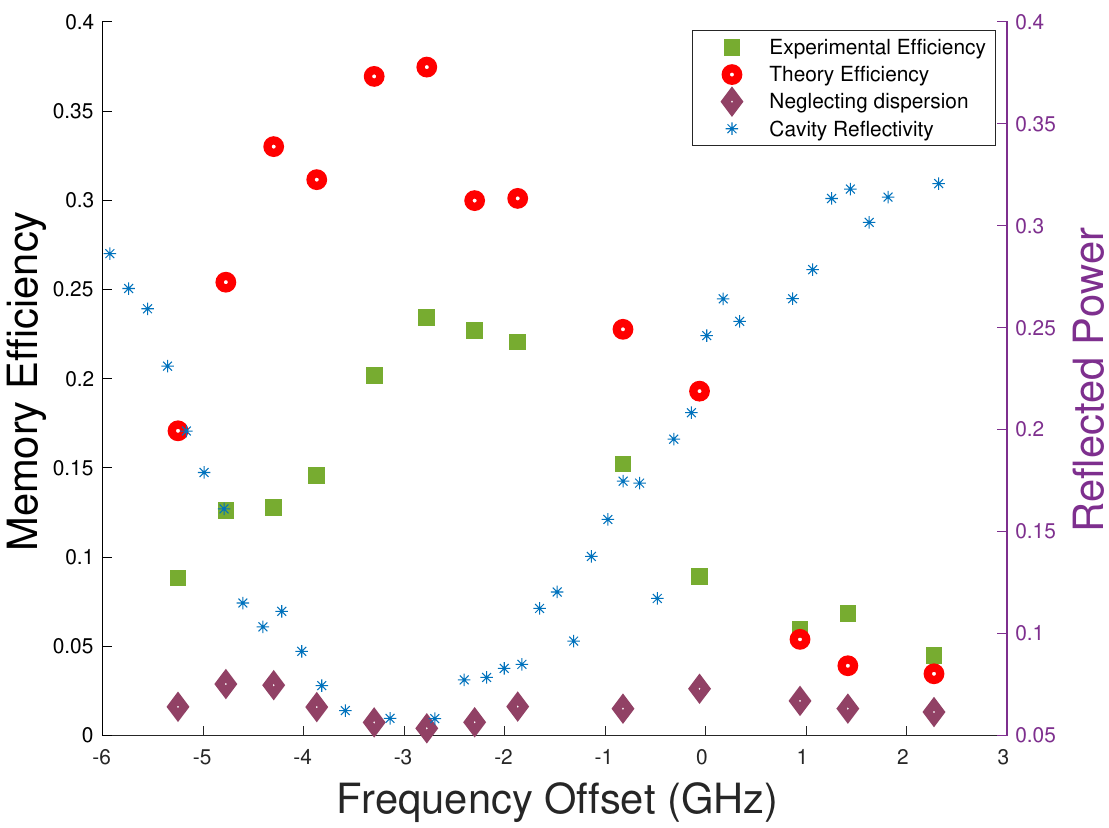}
\caption{The measured experimental efficiency for the created combs and the theoretical efficiency obtained from the model.}  
\label{efficiency}
\end{figure}

We now calculate the efficiencies of the created cavity-enhanced AFC quantum memories numerically taking the following steps. First, we take the Fourier transform (FT) of the input pulse sent for storage to the memory. Then by multiplying the input pulse in the frequency domain by the response function of the cavity (which contains the phase information), the output from the cavity AFC quantum memory is obtained in the frequency domain. Finally, the numerical inverse Fourier transform (FT) results in the output in the time domain. Calculating the square of the absolute value of the output to obtain the output intensity and then plotting the output intensity with respect to time one can obtain the cavity-enhanced AFC quantum memory efficiency by calculating the area under the first echo, i.e., the pulse that occurs after the storage time of the memory, and dividing it by the integrated input intensity. The obtained value is the calculated numerical efficiency of the cavity-enhanced AFC quantum memory. The numerical calculation is performed by coding in Matlab and using the Fast Fourier transform (FFT) algorithm. 
\begin{figure}[h!]
\centering
\includegraphics[scale=0.17]{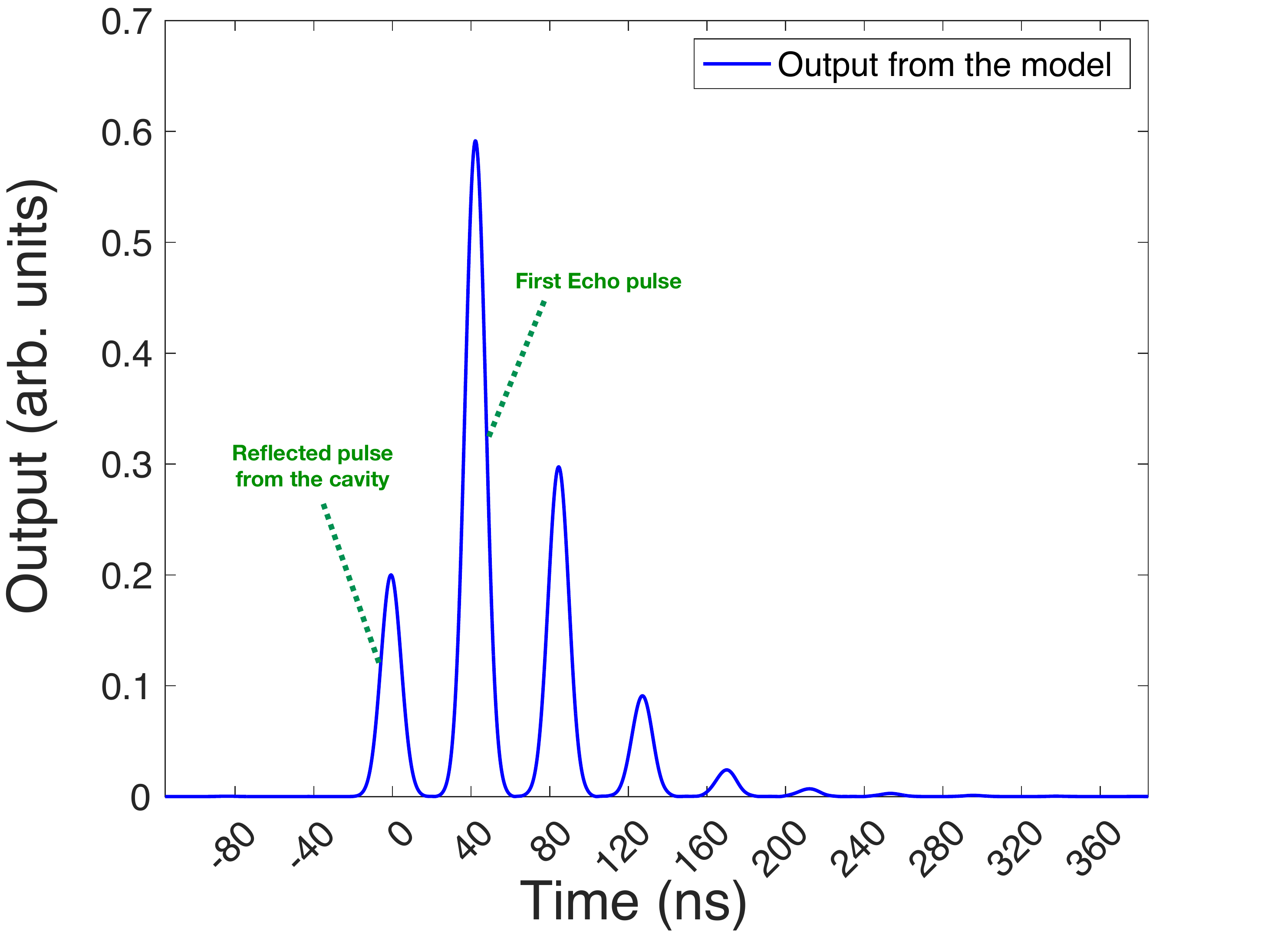}
\caption{The output pulses obtained theoretically for a comb close to the impedance-matched frequency. Memory storage time of $42 \,\mathrm{ns}$ is visible which is the difference between the occurrence time of the reflected part of the input pulse (not being absorbed and stored) and the first recalled (echo) pulse.}  
\label{echo}
\end{figure}
\\\indent Figure \ref{echo} shows the obtained output pulses with respect to time from the model for a cavity-enhanced AFC quantum memory created near the impedance-matched frequency. In the obtained output we see several pulses where the first one is part of the input pulse reflected from the cavity, and the other remaining pulses are the echo pulses in a decreasing order. 
\\\indent The calculated efficiencies for all the combs created at different detunings across the cavity features are shown in figure \ref{efficiency} along with the measured experimental memory efficiencies. Based on the experimental data the maximum echo emission (efficiency) occurs for the AFC created close to the impedance-matched frequency. The obtained efficiency trend in the model is the same as the experiment, with the maximum experimental efficiency for the comb created near the impedance-matched frequency. 
\\\indent The model still doesn't capture all features of the experimental reflectivity. For example, there is an asymmetry over the comb bandwidth in the fitted curve in figure \ref{combsall}\,\textbf{c}, which is not exactly the same as the experiment. Also, for the comb created close to the impedance-matched point (figure \ref{combsall}\,\textbf{b}) we don't perfectly capture the features at the top of the experimental curve. Indeed, for different combs, these features appear differently. This leads to an inaccurate estimation of the combs' features and likely to an overestimation of the calculated memory efficiencies. These facts suggest that although the skewness in the data may not seem that important at first look, it is important for being able to predict the combs' features precisely, and, so, to predict efficiencies close to the experimental values. 

\section{Comparison to a theoretical model without dispersion}
\label{comparison}
It is worth investigating the effect of neglecting dispersion due to the atomic absorption, leading to a constant real refractive index
$n$ in equation (2) for the $\mathrm{YAG}$ crystal \cite{zelmon1998refractive,hrabovsky2021optical}. One can do the process of fitting equation \eqref{2} to the experimental data for the cavity reflectivity without and with comb the same as \ref{ref nocomb} and \ref{ref comb} and extract the comb parameters as the fitting parameters. Doing so the skewness at the bottom of the experimental reflected power data for the combs created far from the impedance matched frequency (figure \ref{combsall}\,\textbf{a}), and in an intermediate regime (figure \ref{combsall}\,\textbf{c}), and also the shapes at the top of the comb created close to the impedance matched point (figure \ref{combsall}\,\textbf{b}) are not captured in the fitted curves. 
\\\indent Talking about the obtained fitting parameters although for each comb the obtained values of $\Delta$, and $d_0$ remain relatively the same as before and there is a small change in the obtained value of $\tilde{\gamma}$, considering the pumping procedure (section \ref{expr}) there is a noticeable misestimation in the obtained $d_{c}$ values of the created combs. Our model lets us obtain more reasonable $d_{c}$ values for the created combs. 
\\\indent Employing the same steps as \ref{mem efficiency}, we can calculate the efficiencies of the created AFCs at different detunings across the cavity profile. Figure \ref{efficiency} shows that the calculated efficiency for all of the created combs is lower than the experimental results. Far from the cavity resonance there is only slight disagreement between the theoretical and experimental efficiencies, but as one gets closer to the cavity resonance the disagreement becomes much more prominent. This suggests that dispersion becomes more important for the combs created closer to the impedance-matched frequency (cavity resonance). 
\\\indent Note that the underestimation of the efficiency in the absence of dispersion is not purely due to to the misestimation of $d_{c}$. Using the refractive index n as a fitting parameter, it is possible to obtain reasonable values for $d_{c}$ even in the absence of dispersion, but the predicted efficiencies are still much too low. The underestimation of the efficiency is also due to the violation of causality in the absence of dispersion, which leads to an unphysical echo occurring before the input pulse associated with a reduction of the physical echos.

\section{conclusion and outlook}
\label{conclusion}
In summary, we have developed a model for cavity-enhanced AFC quantum memory, which helps to address the challenges of the measurement of comb properties inside the cavity. Employing the experimental cavity reflectivity, our model allows us to estimate the comb properties and the efficiency of the cavity-enhanced AFC quantum memory with a background absorption and arbitrary bandwidth created at any frequency offset with respect to the cavity resonance. Using the model, one obtains realistic values for the optical depth, as well as the correct trend for the efficiencies (they are maximum near resonance). Furthermore, the predicted values are in semi-quantitative agreement with the experimental ones. For comparison, we have also shown that a model without including dispersion completely fails to predict the memory efficiencies. Our results confirm the important role of dispersion effects in successfully modeling cavity-enhanced AFC quantum memories.  
\\\indent The model still does not capture all the features, which may be related to the fact that predicted efficiencies are somewhat higher than those achieved in experiments. This discrepancy is likely related to a number of points. First, the shape of the individual created teeth is not a perfect match to that of the model; they are closer to a rectangular shape rather than the assumed Gaussian shape. Second, residual reflectivity from the lenses causes some additional modulation on the transmitted intensity - due to a cavity being formed between the lens and the crystal - which affects the obtained parameters for the cavity. Third, averaging the experimental runs for each output trace can influence the experimental efficiency if some of the contributing runs featured lower efficiencies, e.g. from occasional laser instability. In sum, we anticipate that this work can further inform the next generation of cavity-enhanced quantum memories leading to more rapid progress towards their application in quantum networks.

\begin{acknowledgments}
We wish to thank Roohollah Ghobadi, and Erhan Saglamyurek for useful discussions. This work was supported by the Alberta Major Innovation Fund (MIF), the High-Throughput Secure Networks (HTSN) Program of the National Research Council (NRC), Canada Foundation for Innovation (CFI), Alberta Innovates, and the Natural Sciences and Engineering Research Council (NSERC). 
\end{acknowledgments}

%\newpage

\bibliography{References}

\end{document}